\documentclass[showpacs, oneside, twocolumn, amsmath, amssymb, prl, nofootinbib, superscriptaddress,aps]{revtex4-1}
\usepackage{amsfonts,amsmath,amssymb,array,bm,bbm,cases,color,dcolumn,graphicx,mathrsfs,microtype,subfigure,xcolor}
\usepackage{appendix}
\usepackage{float}
\usepackage[colorlinks,linktocpage,linkcolor=cyan,citecolor=cyan]{hyperref}
\definecolor{lightgray}{gray}{0.9}
\definecolor{Amber}{rgb}{1.0, 0.75, 0.0}
\definecolor{blizzardblue}{rgb}{0.67, 0.9, 0.93}

\def\l{\left}
\def\r{\right}
\def\ddd{\mathrm{d}}
\def\be{\begin{equation}}
\def\ee{\end{equation}}
\def\ba{\begin{eqnarray}}
\def\ea{\end{eqnarray}}
\def\bl#1\el{\begin{align}#1\end{align}}

\begin{document}

\title{Binary Supermassive Black Holes Orbiting Dark Matter Solitons:  \\
From the Dual AGN in UGC4211 to NanoHertz Gravitational Waves}

\author{Tom Broadhurst}
\email{tom.j.broadhurst@gmail.com}
\affiliation{University of the Basque Country UPV/EHU, Department of Theoretical Physics, Bilbao, E-48080, Spain}
\affiliation{DIPC, Basque Country UPV/EHU, San Sebastian, E-48080, Spain}
\affiliation{Ikerbasque, Basque Foundation for Science, Bilbao, E-48011, Spain}

\author{Chao Chen}
\email{iascchao@ust.hk}
\affiliation{Jockey Club Institute for Advanced Study, The Hong Kong University of Science and Technology, Hong Kong S.A.R., P.R.China}

\author{Tao Liu}
\email{taoliu@ust.hk}
\affiliation{Department of Physics, The Hong Kong University of Science and Technology, Hong Kong S.A.R., P.R.China}
\affiliation{Jockey Club Institute for Advanced Study, The Hong Kong University of Science and Technology, Hong Kong S.A.R., P.R.China}

\author{Kai-Feng Zheng}
\email{kzhengae@connect.ust.hk}
\affiliation{Department of Physics, The Hong Kong University of Science and Technology, Hong Kong S.A.R., P.R.China}
\affiliation{Jockey Club Institute for Advanced Study, The Hong Kong University of Science and Technology, Hong Kong S.A.R., P.R.China}

\begin{abstract}
	\noindent

We explore orbital implications of the Supermassive Black Hole (SMBH) binary in UGC4211 for the energy spectrum of stochastic gravitational wave background (SGWB), measured with pulsar timing. The SMBH binary in UGC4211 has a projected separation of $\sim 230\,$pc and relative velocity of $\sim 150\,$km/s along the line of sight. It orbits with a disk of gas and stars, with a total mass $\sim 1.7 \times 10^9 M_\odot$ that is several times larger than the combined SMBHs plus the observed gas and stars. The unseen mass can be naturally explained by a soliton of wave dark matter present within the SMBH orbit. Such a scenario is encouraging as during galaxy merger, the two precursor galactic solitons are expected to combine to generate a new soliton and hence bind the two initial SMBHs efficiently. Generalizing this scenario to the cosmological population of SMBH binaries, we show the SGWB spectrum produced by late-stage inspiraling is modified preferentially at low frequency by the presence of soliton. Finally, we demonstrate that the NANOGrav and EPTA data can be well-fit in this scenario, favoring $\{m_a, f_a\} \sim \{10^{-21.7} {\rm eV}, 10^{15.5} {\rm GeV}\}$ and $\{10^{-20.5} {\rm eV}, 10^{16.8} {\rm GeV}\}$ respectively when the UGC4211 data and the constraints from dwarf galaxies are also combined.

\end{abstract}
	
\maketitle

\section{Introduction}

With the advent of nanoHertz (nHz) gravitational wave (GW) detection, using pulsar timing array (PTA) (NANOGrav~\cite{NANOGrav:2023gor}, PPTA~\cite{Reardon:2023gzh}, CPTA~\cite{Xu:2023wog}, EPTA~\cite{EPTA:2023fyk}, etc.), we can examine the evolution of 
supermassive black hole (SMBH) binaries anticipated to dominate the stochastic GW background (SGWB) at nHz frequencies~\cite{NANOGrav:2023hvm, NANOGrav:2023hfp, EPTA:2023xxk, Ghoshal:2023fhh, Ellis:2023dgf, Shen:2023pan} and new physics as well~\cite{Franciolini:2023wjm,Vagnozzi:2023lwo,Niu:2023bsr,Lu:2023mcz,Figueroa:2023zhu,Unal:2023srk,Li:2023bxy,Konoplya:2023fmh, Ghosh:2023aum,Borah:2023sbc,Datta:2023vbs,Murai:2023gkv,Cai:2023dls,Lazarides:2023ksx,Abe:2023yrw,Bian:2023dnv,Lazarides:2023ksx,Zu:2023olm,Han:2023olf,Guo:2023hyp,Ellis:2023tsl,Fujikura:2023lkn,Li:2023yaj,Megias:2023kiy,Bai:2023cqj,Kitajima:2023vre}. The SMBH binaries are understood to be relics of galaxy mergers which have resulted in the  hierarchical growth of structure in the Universe. Supporting this interpretation, examples of binary SMBHs have been uncovered within luminous galaxies in the local universe, including the recently reported dual active
galactic nuclei (AGN) at $z=0.03474$, in UGC4211, a late-stage major galaxy merger where two SMBHs are found to be separated by $\sim 230\,$pc with a mass of $\sim 10^8M_\odot$~\cite{Koss:2023bvr}. Associated with this dual AGN is a rotating disk of stars and gas sharing a common position angle, where their velocities are measured to be $\sim 200 \pm 50\,$km/s, with a maximum radial extension $\sim 180 \pm 35\,$pc. This implies a total mass $\sim 1.7\times 10^9M_\odot$, several times larger than the AGN's, gas and stars combined. Given that UGC4211 was found within a relatively small, volume-limited sample of nearby hard X-ray detected AGN~\cite{Koss:2018ctd}, the inferred cosmological SMBH merger rate may be surprisingly high. 

This newly resolved SMBH binary in UGC4211~\cite{Koss:2023bvr} raises possible new implications for the ingredients required in calculating the nHz SGWB. As mentioned above, there is the possibility of substantial unseen mass responsible for the orbiting gas and stars in this binary. This observation motivates us to examine the potential role of ``Wave'' Dark Matter (WDM)~\cite{Hu:2000ke, Hui:2016ltb} (for a review, see~\cite{Hui:2021tkt, Ferreira:2020fam, Niemeyer:2019aqm}), one of the most important classes of DM candidates, in the formation and evolution of SMBH binaries. In this context,  the unseen mass in UGC4211 could be naturally explained as a WDM soliton, a standing wave of ultralight bosons such as axions or axion-like particles~\cite{Schive:2014dra}, formed from galaxy merger and hence enclosed by this binary.

The UGC4211 case could be generalized to a cosmological population of SMBH binaries. The standard energy spectrum of SGWB from the inspiraling of binary SMBHs then gets modified, due to orbital effects of the central soliton on the binaries, yielding a potential  enhancement at the spectrum low-frequency end as a  smoking-gun signature. The PTA experiments are thus able to provide insights into the WDM properties such as mass and coupling which have determined the soliton profile.

In this Letter, we first model the dual AGN in UGC4211 as  binary SMBHs orbiting a WDM soliton. The soliton as an extra gravitational source yields a shift to the AGN angular velocity. We then calculate the modified SGWB spectrum from such inspralling SMBH binaries, with the local environmental effects combined, and fit it to the recently released NANOGrav~\cite{NANOGrav:2023gor} and EPTA~\cite{EPTA:2023fyk} datasets. Finally we analyze the constraints on the WDM theory by incorporating additionally the UGC4211 dataset and also the existing constraints from dwarf galaxies~\cite{Pozo:2023zmx}.

\section{WDM Soliton Profile}

In the non-relativistic limit, the WDM profile can be described by a classical complex wavefunction $\psi(t, \mathbf{x})$. Its dynamics is  governed by the Schr\"{o}dinger-Poisson equations:   
\bl \label{eq:sp1}
i {\partial \psi\over \partial t} &= \l( - {\nabla^2 \over 2  m_a} +  m_a \Phi_a +  g |\psi|^2 \r) \psi ~,
\\ \label{eq:sp2}
\nabla^2 \Phi_a &= 4 \pi G  m_a \l( |\psi|^2 - n_0 \r) ~,
\el
where $\nabla^2 \equiv \delta^{ij} \partial_i \partial_j$ is spatial Laplacian operator, $\Phi_a$ is gravitational potential generated by the WDM itself,  $n_0$ is background WDM number density~\cite{Schive:2014dra}, and $g = - { 1 \over 8 f_a^2 }$ arises from the small-field expansion of axion-like potential $V(\phi) = {1\over2} m_a^2 \phi^2 - {1\over 4!} {m_a^2 \over f_a^2} \phi^4 + \cdots$. Here $\phi(t, \mathbf{x}) = { 1 \over \sqrt{ 2 m_a } } \l[ \psi(t, \mathbf{x}) e^{- i m_a t} + \psi^*(t, \mathbf{x}) e^{ i m_a t} \r]$ is real and subjects to an attractive self-interaction. 

The self-gravitation leads to the formation of galactic WDM halo with a central solitonic core~\cite{Schive:2014dra}. This soliton has a spherically symmetric profile~\footnote{The central part of soliton could oscillate in time with an order-unity amplitude~\cite{Veltmaat:2018dfz,Niemeyer:2019aqm}. 
Also, the soliton may randomly walk around the central region of the WDM halo~\cite{Schive:2019rrw}. Both of them arise from the wave interference~\cite{Li:2020ryg}. We will not consider these effects in this study as they tend to be averaged out over time.}~\cite{Schiappacasse:2017ham}: 
\be \label{eq:ansatz}
\psi(t, \mathbf{x}) = \chi(r) e^{-i \omega t} ~,
\ee
with $r = |\mathbf{x}|$ and $\omega$ being a chemical potential.  
$\chi(r)$ describes the localized spatial distribution of soliton and can be approximately described by 
\be \label{eq:prof}
\chi(r) \simeq A ~\text{sech}\l({r\over R_s}\r) \, .
\ee
Here $A$ and $R_s$ are the soliton amplitude and core radius, respectively. Both of them are determined by the axion mass $m_a$ and its decay constant $f_a$.  
For a stable soliton, where the gravitation dominates over the attractive WDM self-interaction, we have~\cite{Schiappacasse:2017ham}
\bl
\label{sol:M}
M_s &\sim 10^{2} M_\odot \l( { m_a \over 10^{-20} ~\text{eV} } \r)^{-1} \l( {f_a \over 10^{10} \text{GeV}} \r) ~,
\\ \label{sol:rc}
R_s &\sim 10^5 \text{pc} \l( { m_a \over 10^{-20} ~\text{eV} } \r)^{-1} \l( {f_a \over 10^{10} \text{GeV}} \r)^{-1} ~.
\el
Here $M_s = \int_{0}^{\infty} m_a |\psi(t,r)|^2 \ddd^3r = {1\over3} \pi^3 R_s^3 m_a A^2$ is the total mass of soliton. The soliton density then drops to one half of its peak value at $r=R_s$.

\section{UGC4211 and Binary SMBHs}

Surprisingly, if the binary SMBHs in UGC4211 form a circular orbit with the projected separation as its diameter, the gravity between them can support a stable velocity $\sim 40\,$km/s, about one quarter of their relative velocity along the line of sight only. This implies that up to an observation uncertainty a ``hidden'' mass might be present within the binary orbit, in excess of the two SMBHs plus associated gas and stellar disc combined. Yet, the two SMBHs share the same position angle, which indicates that they rotate in common with the gas and stellar disk. Their spatial distance and orbital velocity are thus bigger than the projected separation and the velocity along the line of sight, respectively. A more accurate estimate therefore should be based on the full-extent observation of the stars and gas. With the maximal rotation speed $\sim  200$ km/s at a radius $\sim 180\,$pc for the gas, one can find a total mass $\sim 1.7\times 10^{9}M_\odot$ within this radius.

One possible explanation to this observation is that the binary SMBHs in  UGC4211 are orbiting a giant WDM soliton, where the required mass is ``hidden''. Such a system could be common in the WDM scenario as a relic of galaxy mergers. The SMBHs are usually hosted by the DM halo at galactic center. As two galaxies merge, the solitons tend to merge into a new core faster - the ejected matter takes away a large portion of energy and angular moment from the original solitons, resulting the formation of a SMBH binary.  Simulations on the isolated two-soliton merger have been performed in literatures (see, e.g.,~\cite{Guzman:2004wj, Schwabe:2016rze, Edwards:2018ccc}). As indicated in~\cite{Schwabe:2016rze}, as much as $\sim 30\%$ of the initial total mass could be ejected by such gravitational cooling.

\begin{figure}[ht]
	\centering
	\includegraphics[width=0.36\textheight]{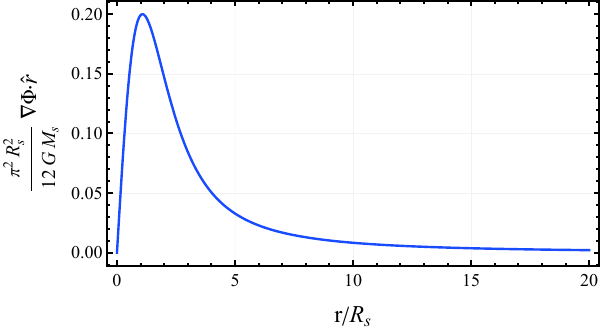}
	\caption{Gravitational force per unit SMBH mass provided by a WDM soliton with the total mass $M_s$ and core radius $R_s$. }
	\label{fig:gf}
\end{figure}

We simply model this system as a binary where the two SMBHs have mass $m_1 = m_2 = M_{\rm BH}$, orbital velocity $v_{\rm orb}$ and spatial separation $r_1 + r_2 = 2 r$. Here $r$ refers to the distance from the soliton center to the SMBHs. The dynamics of this binary is then described by
\be \label{eq:circle}
{v_{\rm orb}^2 \over r} = {G M_\text{BH} \over 4r^2} + \nabla\Phi_a \left(r\right) \cdot \mathbf{\hat{r}}  ~ ,
\ee
where 
\be \label{eq:force}
\begin{aligned}
\nabla\Phi_a(r) \cdot \mathbf{\hat{r}}
&= {12 G M_s \over \pi^2 r^2} \int_0^{\frac{r}{R_s}} x^2 \text{sech}^2(x) \ddd x 
\\&= \Bigg\{
\begin{matrix}
{4 G M_s \over \pi^2 R_s^3} r ~, &r \ll R_s ~,
\\ 
{G M_s \over r^2} ~, &r\gg R_s ~,
\end{matrix}
\end{aligned}
\ee
denotes the gravitational force provided by the soliton. We show this force as a function of $r/R_s$ in Fig.~\ref{fig:gf}. For the inner region where $r \ll R_s$, the density of soliton core is approximately flat and hence $\nabla\Phi(r) \cdot \mathbf{\hat{r}}$ scales linearly with $r$. For the ourter region where $r \gg R_s$, however, the soliton core behaves like a point mass whose gravitational force follows the inverse square scaling.

\section{SGWB from Binary SMBHs Orbiting a Soliton}

UGC4211 could be just one of the enormous amount of binary SMBHs orbiting a WDM soliton formed in cosmic history. As these binaries evolve to a smaller separation, the nano-hertz GWs can be produced, which are expected to be detected with the PTAs as a stochastic background. This stage corresponds to a binary with a separation smaller than the soliton size or inside the soliton. Then the kinematics of its two SMBHs in radial direction is described by  
\begin{equation}\label{eqbinary}
\begin{aligned}
\omega_\text{orb}^2 r_1 m_1 &=\frac{Gm_1m_2}{R^2} +\frac{4GM_s}{\pi^2R_s^3}m_1r_1 ~,
\\
\omega_\text{orb}^2 r_2 m_2 &=\frac{Gm_1m_2}{R^2} +\frac{4GM_s}{\pi^2R_s^3}m_2r_2 ~,
\end{aligned}
\end{equation}
where $\omega_\text{orb}$ represents the orbital angular velocity of the system, and $R = r_1+r_2$ denotes the separation between the two SMBHs.  
$\omega_\text{orb}$ is solved to be 
\begin{equation}\label{eq:orb_freq}
\omega_\text{orb}^2 = {G(m_1+m_2) \over R^3} + {4 G M_s \over \pi^2 R_s^3} ~,
\end{equation}
where ${4 G M_s \over \pi^2 R_s^3}$ decodes a new environmental effect induced by the soliton. The orbital binding energy is then  
\begin{eqnarray}  \label{eq:Eorb}
E_\text{orb} &=&  {1\over2}m_1 \omega_\text{orb}^2r_1^2+ {1\over2}m_2 \omega_\text{orb}^2r_2^2- {G m_1 m_2 \over R} \nonumber \\ && + m_1\Phi_a(r_1)+m_2\Phi_a(r_2) ~.
\end{eqnarray}
Here we do not require $m_1=m_2$ or $r_1=r_2$, but assume the soliton to be spherically symmetric with respect to the binary mass center. 

The energy spectrum of SGWB in a homogeneous and isotropic Universe can be calculated with~\cite{Phinney:2001di}   
\be \label{eq:omegaGW_def}
\Omega_\text{gw}(f) \equiv {1 \over \rho_c} {\ddd \mathcal{E}_\text{gw} \over \ddd \ln f}
= - {1 \over \rho_c} \int_0^\infty \ddd z {N(z) \over 1+z} {\ddd E_\text{orb} \over \ddd \ln f_r} \Big|_{f_r=f(1+z)} ~. 
\ee
Here $\mathcal{E}_\text{gw}$ is energy density of GWs at present, $\rho_c =3H_0^2/(8\pi G)$ is cosmic critical density defined with the current Hubble constant $H_0$, $1/(1+z)$ denotes the effect of GW redshift, $N(z) \ddd z$ is GW event count per unit comoving volume in a redshift interval $(z, z+ \ddd z)$, $f_r$ and $f = f_r/(1+z)$ are GW frequencies produced originally and observed currently, and $- \ddd E_\text{orb} / \ddd \ln f_r$, determined by the underlying physics, describes frequency energy spectrum of GWs radiated in individual events. As the frequency of GWs emitted by a binary is twice the frequency of its orbital motion~\cite{Maggiore:2007ulw}, namely $f_r = \omega_\text{orb}/\pi$, we obtain
\be \label{eq:single_E}
 {\ddd E_\text{gw} \over \ddd \ln f_r} = - {\ddd E_\text{orb} \over \ddd \ln f_r}
= {\pi^{2/3} \over 3 G} (G\mathcal{M})^{5/3} f_r^{2/3} { 1 + 3 (f_s/f_r)^2 \over \l[ 1 - (f_s/f_r)^2 \r]^{5/3} } ~.
\ee
Here $\mathcal{M} = (m_1m_2)^{3/5} (m_1+m_2)^{-1/5}$ denotes chirp mass.  
\be \label{eq:fsol}
f_\text{s} \equiv \sqrt{ {4 G M_s \over \pi^4 R_s^3} }
\sim 1.6\times 10^{-22} ~\text{Hz} \l( { m_a \over 10^{-20} ~\text{eV} } \r) \l( {f_a \over 10^{10} \text{GeV}} \r)^2 
\ee
measures frequency shift of the radiated GWs for a binary inside the soliton. This frequency shift is originally caused by the soliton gravitational attraction (see Eq.~\eqref{eq:orb_freq}). For $f_s \to 0$,
Eq.~\eqref{eq:single_E} is reduced to a standard format~\cite{Phinney:2001di,Maggiore:2007ulw} where no soliton plays a role. By plugging Eq. \eqref{eq:single_E} into Eq. \eqref{eq:omegaGW_def}, we obtain the SGWB energy spectrum (the power of $f/f_{\rm ref}$ is often referred to as $2 + 2\alpha$)
\be \label{eq:gw}
\Omega_\text{gw}(f) = A_\text{gw}^2 \l( {f\over f_{\rm ref}} \r)^{{2\over3}} { 1 + 3 (f_\text{s}/f)^2 \over \l[ 1 - (f_\text{s}/f)^2 \r]^{5/3} } ~,
\ee
where $f_{\rm ref} \equiv 1 ~\text{yr}^{-1}$ is a reference frequency, and 
\begin{eqnarray} 
A_\text{gw} = f_{\rm ref}^{1/3} \sqrt{ {8 \pi^{5/3} \over 9 H_0^2} (G\mathcal{M})^{5/3} N_0 \l\langle (1+z)^{-1/3} \r\rangle } \label{Agw}
\end{eqnarray}
is the amplitude of spectrum. In Eq.~(\ref{Agw}), $N_0 = \int_{0}^{\infty} N(z) \ddd z$ is the comoving number density of merger events by today and $\l\langle (1+z)^{-1/3} \r\rangle = {1\over N_0} \int_{z_\text{min}}^{z_\text{max}} {N(z) \over (1+z)^{1/3}} \ddd z$ is a redshift factor with $z_\text{min} = \max[0, f_\text{min}/f-1]$ and $z_\text{max} = f_\text{max}/f-1$~\cite{Phinney:2001di}. $f_\text{max}$ and $f_\text{min}$ are set by the binary separation at its birth and the frequency at which it comes into Roche lobe contact, respectively~\cite{Phinney:2001di}. As the galaxy merger is expected to occur late (typically at a redshift $z \lesssim 1$~\cite{McWilliams:2012an}), while deriving Eq.~(\ref{eq:gw}) we have taken an approximation of $f_s/f_r \approx f_s/f$ to simplify the analysis. Notably, the correction factor in Eq.~\eqref{eq:gw} is bigger than unity for $f_s \neq 0$, but reduced to one in the trivial case. As $f_s < f_r \sim f$, the $\Omega_\text{gw}(f)$ receives more positive corrections at a lower frequency. This can be well-explained. Due to the presence of central soliton in the binary, the two SMBHs can stay from each other farther to radiate the GWs of the same frequency, compared to the standard scenario. In this case, the quadrupole moment of the binary SMBHs gets enhanced. But, as the separation between the two SMBHs becomes smaller, where the radiated GWs have a higher frequency, the hidden mass enclosed by these two SMBHs gets reduced. We anticipate this effect to be qualitatively unchanged even if the circular orbital is deformed to an elliptical one.

In detail, the evolution of binary SMBHs is also expected to be subjected to local ``environmental effects", including dynamical friction from viscous drag after the host galaxies merge ($R \sim 10 ~{\rm kpc} - 100~{\rm pc}$)~\cite{Antonini:2011tu, Kelley:2016gse, Dosopoulou:2016hbg}, stellar loss-cone scattering ($R \sim 100 ~{\rm pc} - 0.1~{\rm pc}$), and also viscous circumbinary gas and disk interaction ($R \sim 0.01 ~{\rm pc} - 0.001 ~{\rm pc}$)~\cite{Begelman:1980vb, Frank:1976uy,Milosavljevic:2002bn}. The GW radiation becomes dominant for $R < \mathcal O(0.001)\,$pc. Due to the complexity of these effects, considerable efforts are  dedicated to numerical simulations (see, e.g.,~\cite{Munoz:2018tnj,Moody:2019nes}). Here we make a simplified treatment and consider
\be \label{eq:dEgw_dlnfr}
{\ddd E_{\rm gw} \over \ddd \ln f_r} 
={\ddd E_{\rm gw} \over \ddd t} {\ddd t \over \ddd \ln R}{\ddd \ln R \over \ddd \ln f_r}   \, ,
\ee
where $\frac{\ddd \ln R }{ \ddd \ln f_r}$ reflects the relation between $\omega_{\rm orb}$ and $R$ in the presence of a soliton, $\ddd E_{\rm gw} \over \ddd t$ denotes the GW radiation power of the binary, and $- \frac{\ddd t }{ \ddd \ln R} = - R / \dot{R}$ (denoted as $t_{\rm eff}$ below) characterizes the time scale of its orbital decay. $t_{\rm eff}$ is jointly determined by the GW radiation and the environmental effects, namely as $t_{\rm eff}^{-1} = t_{\rm gw}^{-1} + t_{\rm env}^{-1}$. Here $t_{\rm gw}$ and $t_{\rm env}$ are time scales for orbital decay caused by the GW radiation and environmental effects respectively.  We calculate $t_{\rm gw}$ by applying the quadrupole radiation power 
\be \label{eq:dEgw_dt_0}
{\ddd E_{\rm gw} \over \ddd t} = {32 \over 5} G^{{7\over3}} (\pi \mathcal{M} f_r)^{{10\over3}} \l[ 1 - \l(\frac{f_s}{f_r}\r)^2 \r]^{-{4\over3}} 
\ee
and the solution of Eq.~\eqref{eq:orb_freq}
\be \label{eq:R_fr}
R = \l[ {G (m_1 + m_2) \over \pi^2 (f_r^2 - f_s^2 ) } \r]^{1\over 3}  
\ee
to Eq.~(\ref{eq:dEgw_dlnfr}) (note, Eq.~(\ref{eq:dEgw_dlnfr}) can be applied for the case with no environmental effects also where $t_{\rm eff}$ is reduced to $t_{\rm gw}$), which yields
\be \label{eq:teff_0}
t_{\rm gw} = {5 \over 64} (G\mathcal{M})^{-{5\over3}} (\pi f_r)^{-{8\over3}} \l[1 + 3 \l(\frac{f_s}{f_r}\r)^2 \r] \l[ 1 - \l(\frac{f_s}{f_r}\r)^2 \r]^{{2\over3}} .
\ee
$t_{\rm env}$ is smaller than $t_{\rm gw}$ for the binary phase of emitting nHz GWs, and at the leading order we have $t_{\rm env} \approx C R^\beta$~\cite{Haiman:2009te,Kocsis:2010xa,Ellis:2023dgf}. 
As $\ddd E_{\rm gw} \over \ddd t$ and $\frac{ \ddd \ln f_r}{\ddd \ln R }$ for a given $R$ value are not sensitive to the environmental effects~\cite{Kocsis:2010xa,Kelley:2016gse, Ellis:2023dgf}, the overall $\Omega_{\rm gw}(f)$ can be approximately calculated by multiplying Eq.~(\ref{eq:gw}) with an factor $\frac{t_{\rm eff}}{t_{\rm gw}}$.  
Finally we have the SGWB energy spectrum 
\bl \label{eq:omega_gw_res}
\Omega_{\rm gw}(f)
&\approx A'_{\rm gw}{}^{2} \l( {f\over f_{\rm ref}} \r)^{10-2 \beta \over 3} \\
&\times \l[ 1 - \l(\frac{f_s}{f}\r)^2 \r]^{-(7+\beta)\over 3} \l[ 1-\l(f\over f_{\rm gw}\r)^{8-2\beta\over 3}\r]~, \nonumber 
\el
where $A'_{\rm gw} = A'_{\rm gw} (A_{\rm gw},C,\beta)$ is a counterpart of $A_{\rm gw}$ in Eq.~(\ref{eq:gw}) with the environmental effects and 
\be
f_{\rm gw}= {1\over\pi} \l[\frac{5}{64 C } \l(G\mathcal{M}\r)^{-5 \over 3} [G(m_1 + m_2)]^{-\beta \over 3} \r]^{3\over 2\beta - 8} \ ,
\ee
due to $\frac{t_{\rm gw}}{t_{\rm env}}|_{f_r=f_{\rm gw}} \approx 1$, sets up a maximal frequency for the GWs radiated before the last binary phase where $t_{\rm gw} < t_{\rm env}$. The last two factors in Eq.~(\ref{eq:omega_gw_res}) result from the soliton and environmental effects and influence more the spectrum at the low- and high-frequency ends, respectively. In the limit of $f_s \to 0$, Eq.~(\ref{eq:omega_gw_res}) is reduced to the case in~\cite{Haiman:2009te,Kocsis:2010xa,Ellis:2023dgf} where no soliton is present.

\section{Data Fitting}

\begin{figure}[ht]
\centering
\includegraphics[width=0.36\textheight]{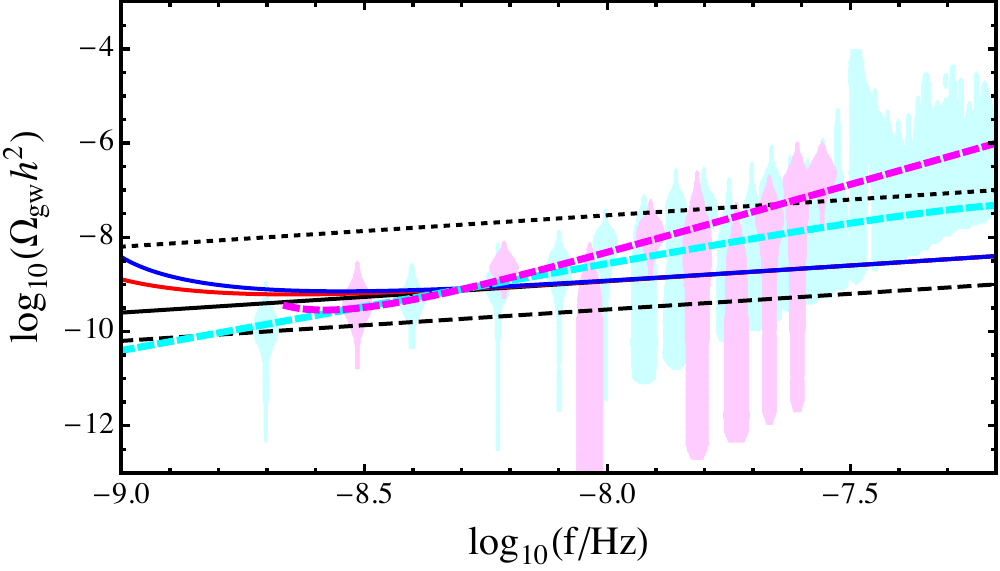}
\caption{Data fitting for the SGWB energy spectrum  generated by the binary SMBHs with a WDM soliton. The black solid curve represents the best-fit of Eq.~(\ref{eq:gw}) to the NANOGrav 15-year dataset~\cite{NANOGrav:2023gor}, with $\{ \log_{10} A_{\rm gw}, \log_{10}(f_s/\text{Hz}) \} = \{-4.1, < -11.6\}$~\footnote{The posterior probability becomes approximately flat as $f_s \to 0$ for the fitting of the NANOGrav 15-year dataset, so we have set up the upper limit at $68\%$ C.L. instead in this direction.}. The black dotted and dashed curves represent a shift from the best-fit point with $\log_{10} A_{\rm gw} = -3.4$ and $-4.4$, respectively, and the red and blue solid curves do so with $\log_{10}(f_s/\text{Hz}) = -9.2$ and $-9.1$ respectively. We present an overall fitting of Eq.~\eqref{eq:omega_gw_res} to the NANOGrav 15-year dataset~\cite{NANOGrav:2023gor} and the EPTA dataset~\cite{EPTA:2023fyk} as dashed cyan and magenta curves, with the best-fit values: $\{ \log_{10}  A'_{\rm gw}, \log_{10}(f_s/\text{Hz}), \beta,  \log_{10}(f_{\rm gw}/\text{Hz}) \}= \{-3.8, < -11.2, 2.2,-6.9\}$ and $\{-3.4, -8.8, 0.6,-6.1 \}$, respectively. 
}
\label{fig:fs}
\end{figure}

We show in Fig.~\ref{fig:fs} the SGWB energy spectrum generated by the binary SMBHs with a WDM soliton, by fitting to the NANOGrav 15-year dataset~\cite{NANOGrav:2023gor} and the EPTA dataset~\cite{EPTA:2023fyk}. In their recent reports, 
NANOGrav~\cite{NANOGrav:2023gor}, PPTA~\cite{Reardon:2023gzh}, CPTA~\cite{Xu:2023wog} and EPTA~\cite{EPTA:2023fyk} have impressingly demonstrated the consistency between the data and the Hellings-Downs curve, strongly supporting the SGWB interpretation of the PTA signals. But, a deviation of $\alpha$ from its prediction $-2/3$ in the standard SMBH binary scenario has been observed. The observed $\Omega_{\rm gw} (f)h^2$ spectrum is steeper, favoring a less negative $\alpha$ value in the power law fitting~\cite{NANOGrav:2023gor}. This discrepancy however can be resolved by the environmental effects on the overall frequency factor in Eq.~(\ref{eq:omega_gw_res}), with $\beta < 4$. As indicated by this figure, the best-fit to the NANOGrav 15-year dataset and the EPTA dataset~\cite{EPTA:2023fyk} favors $\beta =2.2$ and $0.6$, for a frequency range between $\{f_s, f_{\rm gw}\} = \{< 10^{-11.2}, 10^{-6.9}\}\,$Hz and $\{10^{-8.8}, 10^{-6.1}\}\,$Hz respectively. More than that, to better fit the EPTA dataset requires $f_s > 0$, resulting in an apparent enhancement to the spectrum at its low-frequency end. This feature if being confirmed can serve as a smoking-gun signature for the scenario of binary SMBHs orbiting a central soliton. The enhancement could be extended to a frequency as small as $f_s$. For $f < f_s$, the correspondent binary separation becomes larger than the soliton diameter and the two SMBHs thus have not evolved into the soliton yet.

\begin{figure}[ht]
\centering
\includegraphics[width=0.36\textheight]{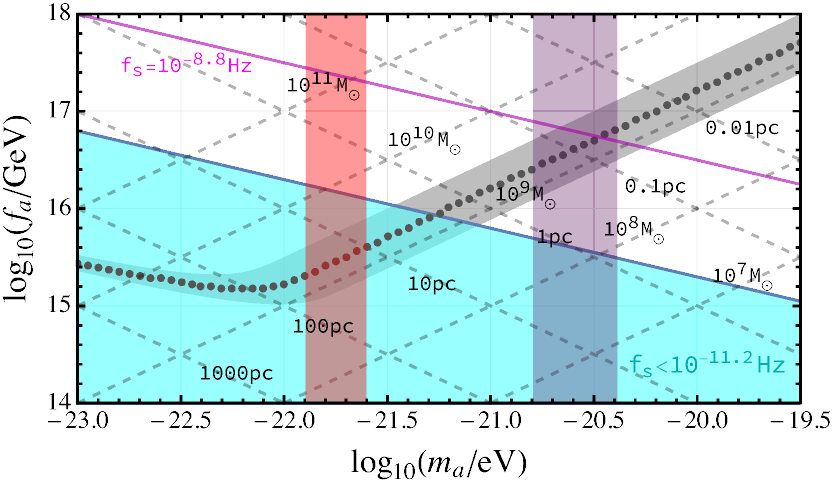}
\caption{Constraints on $m_a$ and $f_a$. The shaded grey band is obtained by fitting the model (see Eq.~\eqref{eq:circle}) to the UGC4211 data~\cite{Koss:2023bvr}. The shaded cyan region and magenta line denote the favored $f_s$ values by the NANOGrav 15-year dataset~\cite{NANOGrav:2023gor} and the EPTA dataset~\cite{EPTA:2023fyk}, respectively. The shaded purple and red regions correspond to the favored $m_a$ ranges by the datasets of ultra faint dwarfs and dwarf Spheroidal galaxies, respectively~\cite{Pozo:2023zmx}. As a reference, we also present total mass $M_s$ and core radius $R_s$ of the soliton as grey dashed contours.
}
\label{fig:mafa}
\end{figure}

Finally, we interpret the NANOGrav~\cite{NANOGrav:2023gor} and EPTA datasets~\cite{EPTA:2023fyk}, together with the UGC4211 dataset~\cite{Koss:2023bvr}, in the $m_a -f_a$ plane in Fig.~\ref{fig:mafa}. The existing constraints from the datasets of dwarf galaxies~\cite{Pozo:2023zmx} are also considered. As shown in this figure, the shaded grey band is separated into two parts by a turning point: $m_a \lesssim$ and $\gtrsim 10^{-22.1}\,$eV, where the UGC4211 binary orbits inside and outside the soliton respectively. As discrepancies exist between the fittings to the NANOGrav and EPTA datasets and also to the datasets of ultra faint dwarfs and dwarf Spheroidal galaxies~\cite{Pozo:2023zmx}, a degeneracy arises in this analysis: the UGC4211 dataset, while combining with the datasets of EPTA and ultra faint dwarfs, favors $\{m_a, f_a\} \sim \{10^{-20.5} {\rm eV}, 10^{16.8} {\rm GeV}\}$, but prefers $\{m_a, f_a\} \sim \{10^{-21.7} {\rm eV}, 10^{15.5} {\rm GeV}\}$ in joint with the datasets of NANOGrav and dwarf Spheroidal galaxies. To resolve this degeneracy requires inputs from future observations.

\section{Summary and Outlook}

In this Letter, we have explored the orbital effects of a central WDM soliton on the SMBH binaries, which can be applied to address the newly resolved dynamics of the dual AGN in UGC4211 and may leave an imprint in the SGWB energy spectrum being actively measured with the PTA. The scenario of central soliton is generic for the WDM theory, formed from galaxy merger and hence enclosed by the binary. As the nHz SGWB has been long anticipated to be dominantly sourced by the SMBH binaries, it is timely to explore the phenomenology of this scenario. Then, by modeling the dynamics of binary SMBHs circularly orbiting a WDM soliton and integrating their contributions over cosmological history, with the local environmental effects combined, we showed that the produced SGWB energy spectrum can well-fit the observational PTA data. Particularly, the EPTA dataset favors an apparent enhancement at the spectrum low-frequency end, distinguishing the soliton scenario from the standard SMBH binaries. We then identified the favored parameter values for the WDM theory, by incorporating additionally the UGC4211 dataset and also the existing constraints from dwarf galaxies. Yet, to fully resolve this scenario requires detailed simulations of the binary orbital evolution with a soliton, together with the merger of two precursor galactic solitons, and refined analysis by including new inputs from future observations. We leave these explorations to an ongoing project~\cite{BCLZ}.

\section*{Acknowledgement}

C.~Chen would like to thank Jie-Wen Chen, Chengjie Fu, Leo WH Fung, Yudong Luo, Hoang Nhan Luu, Xin Ren, Chengfeng Tang, Zi-Qing Xia and Guan-Wen Yuan for useful discussions. This project is supported by the Collaborative Research Fund under Grant No. C6017-20G which is issued by Research Grants Council of Hong Kong S.A.R.

\bibliography{binary}

\end{document}